\documentclass[twoside,11pt,preprint]{article}

\usepackage{dmlr2e}
\usepackage{amsmath}
\usepackage{booktabs}
\newcommand{\blfootnote}[1]{\begingroup\renewcommand\thefootnote{}\footnotetext{#1}\addtocounter{footnote}{-1}\endgroup}

\ShortHeadings{STRATA: Name-and-Geography Race Inference}{Chalavadi, Leitch, and Pastor}
\firstpageno{1}
\def\openreview{\url{https://openreview.net/forum?id=9n0t3JC8p8}}

\begin{document}

\title{STRATA: A Name-and-Geography Race Inference Model for Fair Lending and Housing Equity Applications}

\author{\name Sanjana Chalavadi \email sanjana@aequum.ai \\
       \addr aequumAI LLC
       \AND
       \name Terry Leitch \email terry@aequum.ai \\
       \addr aequumAI LLC
       \AND
       \name Andrei Pastor \email andrei@aequum.ai \\
       \addr aequumAI LLC}

\editor{}

\maketitle
\blfootnote{Submitted to \emph{Data-centric Machine Learning Research} (DMLR).}

\begin{abstract}
Accurate aggregate estimation of race and ethnicity (R\&E) can be important in fair-lending and housing-equity analyses when self-reported demographic data are unavailable, settings in which regulated institutions are nonetheless expected to assess disparities. Existing proxies, including the CFPB-used Bayesian Improved Surname Geocoding (BISG), exhibit socioeconomically correlated misclassification that can cause measured disparities to understate true levels. This paper introduces STRATA (Socioeconomic and Tract-Referenced Attribution for Algorithmic analysis), which integrates character-level name sequences with block-group and census-tract geography via stacked Bidirectional LSTM networks and XGBoost post-filtering, trained on Florida, Georgia, and North Carolina voter registrations and the national PPP loan file. On a matched voter holdout ($N = 981{,}288$) evaluated from a common cohort (with BISG/BIFSG metrics computed over the subsets those methods score), STRATA reduces the White False Positive Rate---the rate at which non-White individuals are misclassified as White---from 28.0\% (BISG) to 17.8\%, the lowest of any evaluated method, at 88.8\% accuracy while returning predictions for all eligible tract-resolved records (100\% conditional coverage). On a geographically national held-out PPP cohort of self-identified small-business borrowers (all 50 states, restricted to a rule-based person-name cohort), STRATA achieves 88.5\% accuracy with a 9.81\% White FPR. STRATA requires only a name and 13 public Census features---unlike corrections dependent on voter-file-only covariates (Argyle \& Barber) or $\sim$203 engineered ACS features (ZRP)---making it deployable where aggregate proxy analysis is legally and methodologically appropriate. These models are appropriate for aggregate population-level analysis and should not be used for individual-level transactional decisions.
\end{abstract}


\section*{Errata: Correction to the Previously Posted Version}

The originally posted version of this paper (arXiv:2504.21259v1) reported that STRATA ``reduces the White FPR from 41.8\% for BISG to 17.8\%, the lowest among all models.'' The 41.8\% BISG figure was drawn from a separate production experiment rather than from the holdout used for the paper's other results, so the comparison mixed evaluation conditions; we retract that claim. This version replaces it with a single, rigorously consistent evaluation in which every model---STRATA, BISG, BIFSG, ZRP, and Argyle \& Barber---is applied to the same initial holdout cohort ($N = 981{,}288$); STRATA, ZRP, and A\&B return predictions for every record, while BISG and BIFSG metrics are computed over the subsets they score, and pairwise statistical tests involving those methods use common support. On that basis, the CFPB-used BISG White False Positive Rate is 28.0\%, and STRATA's is 17.8\%---the lowest of any evaluated method, confirmed on the matched holdout. All results in this version supersede those in the originally posted version (v1) of arXiv:2504.21259.

\section{Introduction}

Accurate race and ethnicity (R\&E) imputation is critical for identifying and addressing disparities in fair lending, housing, healthcare, and social science research \citep{Fiscella2006, Elliott2008}. In the context of mortgage lending, the Home Mortgage Disclosure Act (HMDA) requires institutions to report race and ethnicity of applicants, but applicants are not required to self-identify---resulting in missing race data on a substantial share of records. Regulated institutions may need to assess disparities even when race or ethnicity is not disclosed, creating demand for aggregate proxy methods in appropriate compliance and research settings. The absence of a prescribed proxy standard creates a measurement gap that this paper addresses.

Bayesian Improved Surname Geocoding (BISG), which combines surname and geolocation probabilities using Bayesian inference, is among the most widely used proxy approaches in regulatory and research applications \citep{Elliott2009}. However, BISG combines surname and geographic distributions through a fixed Bayesian factorization that treats neighborhood racial composition as a prior on the individual's race, introducing structured misclassification that distorts measured disparities \citep{Greenwald2024, Xin2026}. The direction of this distortion is not uniform: misclassification mixes group effects in a way that tends to shrink estimated disparities toward the majority group \citep{Xin2026}, and in small-business lending BISG proxy errors have been shown to bias measured approval-gap disparities downward by as much as 43\% \citep{Greenwald2024}. In fair-lending settings where minorities are absorbed into the White majority class, the practical consequence is that institutions relying on BISG may understate their disparity exposure.

This paper introduces STRATA (Socioeconomic and Tract-Referenced Attribution for Algorithmic analysis), a method that integrates geolocational information into stacked Bidirectional Long Short-Term Memory (LSTM) neural networks, post-filtered through an XGBoost ensemble, to address these shortcomings. Both BISG and STRATA use an individual's address to associate the record with neighborhood-level Census information; the distinction is in how that information is used. BISG applies a fixed Bayesian factorization of surname and geographic distributions, whereas STRATA learns a discriminative nonlinear interaction between character-level name representations and block-group and tract covariates, reducing the socioeconomic-status (SES)-correlated misclassification that makes BISG unreliable in gentrifying and mixed-income areas.

We evaluate STRATA using a large Florida, Georgia, and North Carolina voter dataset and validate nationally using the Paycheck Protection Program (PPP) loan file. We compare performance against standalone character-level LSTM \citep{Xie2021}, traditional BISG \citep{Elliott2009}, BIFSG \citep{Voicu2018}, and the Zest Race Predictor (ZRP), an XGBoost-based ensemble \citep{ZestAI}. STRATA offers notable improvements over all baseline methods, achieving strong performance while reducing misclassification bias, particularly the misclassification of non-White individuals as White.

Much of this work was initiated by the authors while researching race and ethnicity prediction for the Social Security Administration (SSA). A companion paper applies STRATA to 2.26 million New York City residential deed transactions (1966--2025) to measure homeownership transfer patterns by race \citep{ChalavadiPastorLeitch2026}.

\section{Existing Research: A Survey of Race and Ethnicity Imputation Methods}

\subsection{Introduction}
When self-reported race and ethnicity data are missing, analysts use predictive models to impute an individual's race/ethnicity based on proxies like name and location. Traditional methods rely on Bayesian probability using surnames and geographies, while newer approaches use machine learning (ML) on names (including neural networks) and hybrid ensembles. This report compares key models: BISG, BIFSG, LSTM-based neural networks (including an LSTM+Geo precursor to STRATA developed by the authors), and ensemble methods (e.g.\ Argyle \& Barber BIFSG post-filtering with random forest \citep{Argyle2023} and related hybrid models combining proxy outputs with tree-based post-filters). We examine each method's performance, required data, strengths, and limitations, drawing on the published literature and the authors' own validation experience.

Drawing on the authors' work on race and ethnicity prediction at the Social Security Administration (SSA), this section documents our observations on the methods used and developed. Each has its uses, alone or in combination, depending on the problem and available resources. Once geolocated information is added to models, prediction quality improves, but care has to be taken with the False Positive Rate (FPR) caused by data imbalance: minimizing errors favors the majority class. We tried numerous schemes including SMOTE and manual minority population magnifications but did not see improvement beyond a name plus geolocated information predictions fed into a boosted decision tree filter. We implemented all via our pipeline process described in the Data section.

\subsection{Bayesian Improved Surname Geocoding (BISG)}
BISG is a proxy method that applies Bayes Theorem to an individual's surname and location to infer race/ethnicity \citep{Argyle2023}. It uses two data sources: (1) a surname frequency list indicating the proportion of people with that last name who are White, Black, Hispanic, Asian/Pacific Islander, etc., and (2) the racial composition of the person's residential area (Census tract or block group) \citep{Argyle2023}. By combining these, BISG produces probabilities for each race. It has become a standard and is implemented in tools like the \textit{wru} R package \citep{ImaiKhanna2016} or the SURGEO Python package \citep{surgeo}.

\textbf{Performance:} BISG is accurate at the aggregate level, but individual-level errors occur. Reported overall accuracy is often in the low-to-mid 80\% range \citep{Argyle2023}. BISG's overall error rate on Florida voter data is 15.1\% (85\% accuracy). BISG almost never predicts the Other category (99\% false negative rate) \citep{Argyle2023}.

\textbf{Limitations:} A key limitation is that BISG assumes surname and location independence conditional on race and relies on static Census distributions. Argyle and Barber (2024) showed that error rates correlate with socioeconomic and demographic factors: racial minorities in affluent neighborhoods are often misclassified as White by BISG \citep{Argyle2023}. Greenwald et al.\ (2024) demonstrated that, in small-business lending, these proxy errors bias measured approval-gap disparities by up to 43\% \citep{Greenwald2024}. Critically, \citet{Xin2026} (a 2026 preprint) show that this distortion is not a problem that higher classification accuracy alone can solve: when proxy race is used as a categorical regressor, the resulting coefficients become confusion-matrix--weighted mixtures of the true group effects, so disparity estimates can be shrunk toward the majority group even when overall accuracy is high. This implies that reducing the \emph{specific} errors that drive majority-class absorption---rather than maximizing aggregate accuracy---is important for reducing one source of bias in disparity measurement.

\subsection{BIFSG -- Adding First Name to BISG}
Bayesian Improved First Name and Surname Geocoding (BIFSG) augments BISG by incorporating the individual's first name as an additional predictor \citep{Voicu2018, ImaiOlivellaRosenman2022}. Incorporating first names yields modest but meaningful gains. In Voicu's validation on 279k mortgage applications, BIFSG achieved a weighted average correlation of 0.84 versus 0.82 for BISG \citep{Voicu2018}. The biggest improvement was for identifying Black individuals; the correlation for the Black cohort rose from 0.71 with BISG to 0.75 with BIFSG \citep{Voicu2018}.

\subsection{LSTM-Based Name Models (Neural Networks)}
Researchers explored deep learning models, particularly Long Short-Term Memory (LSTM) neural networks, to predict ethnicity from names. Sood and Laohaprapanon (2018) pioneered this approach, training a character-based LSTM on 13M Florida voter names and achieving 86--87\% overall accuracy. Xie (2021) introduced the R package \textit{rethnicity} with a bidirectional LSTM achieving an F1-score of 0.73 for Black individuals \citep{Xie2021}---more equitable than BISG (F1 0.30--0.40 for Blacks). In the STRATA architecture, we extend this approach with geolocation integration described in Section~3.

\subsection{raceBERT (Neural Networks)}
Parasurama introduced raceBERT, a Transformer-based model using BERT and RoBERTa \citep{Parasurama2021}. In our own evaluation, raceBERT achieved approximately 80\% accuracy on a one-million-record holdout set. We did not pursue raceBERT further, given its inferior performance relative to the LSTM in our evaluation and the added complexity of integrating geolocation. The character-level BiLSTM was better suited to our setting, modeling names as character sequences at substantially lower computational cost.

\subsection{Hybrids: Geo Models with Boosted Decision Tree Post Filters}
Researchers explored ensemble approaches blending Bayesian proxy outputs with additional features. Argyle and Barber (2024) introduced a two-step model: BISG followed by a random forest classifier. This ensemble reduced misclassification of Black individuals from 33.5\% to 23.1\% \citep{Argyle2023}. In the authors' prior work, an XGBoost ensemble over proxy outputs achieved roughly 88\% accuracy versus about 83\% for individual models. ZRP (Zest AI, 2022) trains an XGBoost classifier using surname, first name, and 167 location-based variables \citep{ZestAIRelease2022}, achieving roughly 85--88\% accuracy.

The Argyle and Barber correction is, to date, the strongest published bias-reduction result against BISG, but it is not directly transferable to lending, deed, and small-business-loan records without suitable alternative variables. Its voter-file-specific covariates are generally unavailable in mortgage, deed, and small-business-loan datasets, limiting direct transfer without appropriate substitute variables. This is a practicality limitation, not a modeling one: the correction may benefit from covariates that are, by construction, unavailable outside the voter file. STRATA instead uses only a name string and public Census geography, both available for any US address regardless of domain, which is what makes the head-to-head comparisons in Section~4 relevant to deployment rather than only to the research literature.

\begin{table}[htbp]
\centering
\caption{Illustrative Performance Metrics of Prior Race/Ethnicity Proxy Methods (from Literature)}
\label{tab:prior_work_summary}
\begin{tabular}{lccccc}
\toprule
Method & Accuracy & Precision & Recall & F$_1$-score & FPR (overall) \\
\midrule
BISG (Surname + Geo) & 0.84--0.85 & 0.84 & 0.84 & 0.80--0.85 & $\approx$0.08 (macro) \\
BIFSG (+First Name) & 0.85--0.86 & 0.85 & 0.85 & 0.82--0.86 & $\approx$0.07 (macro) \\
Argyle \& Barber RF & 0.85--0.88 & 0.87 & 0.87 & 0.63 (macro) & $\approx$0.06 (macro) \\
ZRP (Zest 2022 model) & $\approx$0.86--0.88 & $\approx$0.87 & $\approx$0.86 & $\approx$0.86 & $\approx$0.05 (macro) \\
LSTM (name only) & 0.85 & 0.83 & 0.85 & 0.76 (weighted) & $\approx$0.04 (macro) \\
\bottomrule
\end{tabular}
\parbox{\textwidth}{\small \textit{Note:} Entries are rounded. Each method was evaluated on different datasets. Comparability is restricted.}
\end{table}

\section{Proposed Models}

In this paper, we introduce STRATA, comprising two complementary modeling strategies: sequence modeling with neural networks and nonlinear post-filtering using gradient boosted trees.

\subsection{Data and Feature Engineering}

We derive our primary dataset from publicly available voter registration data from Florida (FL), Georgia (GA), and North Carolina (NC) \citep{sood2017, SBE2022}, combined with the national Paycheck Protection Program (PPP) FOIA loan file (self-identified race, all 50 states and Washington, D.C.), which is used both as additional training data and, on a disjoint held-out split, as a geographically national in-source validation set (held out by record, not by source---the held-out PPP records share the source distribution of the PPP records used in training). All sources contain self-reported race and ethnicity, names, and addresses. The original data were mapped into a unified five-class system: White (Non-Hispanic), Black (Non-Hispanic), Hispanic (any race), Asian/Pacific Islander (Non-Hispanic), and Other. Table~\ref{tab:dataset_stats} reports the class distribution of the voter corpus and of the national PPP validation set.

We note a data-access asymmetry relevant to interpreting the comparisons in Section~4: several of the baselines evaluated here, including Argyle and Barber's random-forest correction and ZRP, were developed and validated against national, all-50-state voter file access (e.g., L2 Political); STRATA's training corpus is restricted to three states' voter registrations plus the PPP file, a narrower geographic footprint than at least some of the baselines it is compared against.

\paragraph{Label Harmonization.}
Merging three state voter files and the PPP file, each with its own race/ethnicity schema, required explicit harmonization, which we document rather than bury. Three corrections were applied before training: (a) a Georgia mapping bug that silently dropped Native American records was fixed, recovering 81{,}042 rows previously excluded from the corpus; (b) Native Hawaiian / Pacific Islander records were standardized into the Asian/Pacific Islander class across all sources, consistent with the analytical aggregation; and (c) \texttt{race\_raw} and \texttt{race\_detail} provenance columns were added so that every harmonized label retains its source coding. Together these corrections make the five-class label definition consistent across all sources. Georgia's ``Other'' category, which absorbs both multiracial individuals and Hispanic-origin leakage relative to Florida and North Carolina, is handled not by a hand-tuned loss weight but by the marginalized cross-entropy of Section~3.2, whose admissible-set mask gives such composite labels a principled training signal.

\begin{table}[htbp]
\centering
\caption{Class distribution of the harmonized modeling corpus (train $+$ held-out, presplit).}
\label{tab:dataset_stats}
\begin{tabular}{lrrrr}
\toprule
 & \multicolumn{2}{c}{Voter (FL+GA+NC)} & \multicolumn{2}{c}{PPP (national)} \\
\cmidrule(lr){2-3}\cmidrule(lr){4-5}
Class & Count & \% & Count & \% \\
\midrule
White NH & 12{,}858{,}847 & 65.2 & 316{,}533 & 37.9 \\
Black NH & 3{,}383{,}994 & 17.2 & 406{,}187 & 48.7 \\
Hispanic & 2{,}471{,}082 & 12.5 & 60{,}263 & 7.2 \\
Asian/PI NH & 383{,}487 & 1.9 & 39{,}641 & 4.7 \\
Other & 628{,}597 & 3.2 & 12{,}011 & 1.4 \\
\midrule
Total & 19{,}726{,}007 & 100 & 834{,}635 & 100 \\
\bottomrule
\end{tabular}
\parbox{\textwidth}{\small \textit{Note:} Counts are the full harmonized, geocoded, race-labeled modeling corpus (before the train/held-out hash split); the stagewise flow to the evaluation cohorts is given in Table~\ref{tab:cohort_flow}. The voter corpus is predominantly White (65.2\%), which is central to interpreting the White FPR: majority-class prevalence is what makes absorption of minorities into the White class the dominant error mode. The PPP records reflect only borrowers who disclosed race (approximately one-quarter of all PPP borrowers) and only those whose address resolved to a census tract; their very different composition provides a geographically national, demographically distinct, in-source held-out evaluation (PPP records also appear in training; the held-out split is disjoint by record, not by source).}
\end{table}

\begin{table}[htbp]
\centering
\caption{Cohort flow from source records to evaluation cohorts. Each stage names its denominator; percentages are of the preceding stage.}
\label{tab:cohort_flow}
\begin{tabular}{lrr}
\toprule
Stage & Voter $N$ & PPP $N$ \\
\midrule
Harmonized, geocoded, race-labeled corpus & 19{,}726{,}007 & 834{,}635 \\
\quad Training partition (hash split) & 18{,}739{,}206 & 709{,}554 \\
\quad Held-out partition (hash split) & 986{,}801 & 125{,}081 \\
\midrule
Evaluation cohort & 981{,}288 & 124{,}914 \\
\quad restriction applied & excl.\ multiracial & person-name filter \\
\quad (removed from held-out) & 5{,}513 (0.6\%) & 167 (0.1\%) \\
\bottomrule
\end{tabular}
\parbox{\textwidth}{\small \textit{Note:} The voter held-out fraction is $986{,}801/19{,}726{,}007 = 5.00\%$ and the PPP held-out fraction $125{,}081/834{,}635 = 14.99\%$, matching the design hash thresholds. The voter evaluation cohort excludes held-out records whose provenance label is multiracial (Section~3.1), so that all methods score the same unambiguous five-class labels. The PPP evaluation cohort restricts to person-name records (Section~4.8, two-stage BusinessType $+$ name filter); all PPP held-out records are tract-resolved by construction, since tract features are required to score. The M-series geographic-transfer models of Table~\ref{tab:loso} predate the M8 corpus reconstruction and use a distinct frozen PPP extract (M1/M2 scored on $n=910{,}762$; M3/M4 on the held-out New York PPP set $n=35{,}974$). Their purpose is limited to comparing training-geography conditions \emph{within} that extract; their absolute values should not be compared directly with the M8 person-name cohort of Table~\ref{tab:ppp-national-results}. Re-fitting M1--M4 on the final person-name cohort is left to future work.}
\end{table}

\paragraph{Geolocation Enrichment.}
To incorporate geographic and socioeconomic context, we resolve voter and PPP addresses to Census block group and tract using an offline, TIGER/Line-based address resolver rather than an external geocoding API, improving both coverage and reproducibility relative to our earlier reliance on the U.S. Census Geocoder API \citep{CensusGeocoder}. Both BISG and STRATA associate each record's address with neighborhood-level Census information; the distinction is in how it is consumed. This design reduces---rather than eliminates---the SES-correlated misclassification bias that causes BISG to misclassify minorities in high-income areas as White: the model still conditions on block-group and tract demographic features, but does so within a discriminative sequence model that learns a nonlinear name--geography interaction rather than through a fixed Bayesian factorization that treats neighborhood composition as a prior on the individual's race. This yields a dataset of $\{$Name, Address, Census race distributions, Income decile$\}$ for neural network training. Holdouts use a deterministic hash split: a 5\% voter holdout (state-stratified, self-reported ground truth) and a 15\% PPP holdout ($N=125{,}081$, self-report concordance). For the voter holdout, multiracial records are reconciled into the Other class at training time but excluded from the evaluation set so that every method is scored on the same unambiguous five-class labels, yielding $N=981{,}288$. All methods are applied to this same initial holdout cohort; STRATA, ZRP, and A\&B return a prediction for every record, whereas BISG and BIFSG are scored only where their surname/first-name reference lists yield a prediction, so their metrics are computed over those scored subsets and pairwise comparisons with them are conducted on common support.

\paragraph{Cohort construction, geocoding provenance, and coverage.} Because STRATA conditions on tract-level features, a record can be scored only if its address resolves to a census tract. Applying our offline TIGER/Line resolver to the $4{,}046{,}652$ PPP records that pass the Stage-1 individual-\texttt{BusinessType} filter (Section~4.8) with a two-token personal-name parse and no business-keyword string, only 0.1\% carry a PO-box address and fewer than 0.1\% fail to resolve to any geography; the binding constraint is geocoding precision. 84.6\% resolve to block-group-or-tract precision (78.4\% block group, 6.3\% tract; components may not sum to the total due to rounding) and can be assigned tract features, while the remaining 15.3\% resolve only to ZIP. This benchmark used the original two-token parse; geocoding reach is a property of address quality and the resolver rather than of the name-token rule, and the evaluation cohort of Section~4.8 relaxes that parse to admit middle names and multiple surnames. We therefore distinguish two coverage notions used throughout this paper. \emph{Conditional model-output coverage} is the fraction of tract-resolved records for which a method returns a prediction; STRATA's is 100\% (unlike BISG and BIFSG, which return no output for names absent from their reference lists even when geography is available). \emph{Eligible-cohort geocoding coverage} is the fraction of records \emph{already satisfying} the individual-\texttt{BusinessType} and person-name-parsing criteria whose addresses resolve to block-group or tract precision; among the $4{,}046{,}652$ records in the benchmark cohort above, this is 84.6\%. This figure characterizes address-resolution reach \emph{within that eligible cohort}; it is \emph{not} the fraction of all raw PPP records that STRATA can score, and it is a historical geocoding benchmark computed on the earlier exact-two-token cohort rather than the measured coverage of the final $N=124{,}914$ evaluation cohort. Every voter- and PPP-holdout metric reported in this paper is computed at 100\% conditional model-output coverage on the applicable evaluation cohort. For the initial PPP holdout ($N=125{,}081$), tract resolution is guaranteed by construction. The final PPP evaluation cohort additionally applies the two-stage person-name restriction described in Section~4.8, yielding $N=124{,}914$; PPP records carry no verified secondary address, so no address-agreement criterion is applied.

\subsection{STRATA Base Model (LSTM+Geo)}

The STRATA base model integrates the sequential characteristics inherent to names (processed with Bidirectional LSTM networks) and demographic data from census tract geolocation. This hybrid sequence-based model significantly enhances prediction accuracy and reduces socioeconomic biases prevalent in existing Bayesian geocoding methods \citep{Elliott2009, Argyle2023}.

\paragraph{Model Architecture and Features.}
\textbf{Input Features:}
\begin{itemize}
\item Character-level embeddings representing sequences of characters in first, middle, and last names.
\item Geolocation attributes (13 public Census features): block-group racial composition (five rates), block-group median-income decile, and block-group population; census-tract racial composition (five rates) and tract median-income decile.
\end{itemize}
\textbf{Neural Network Architecture:}
\begin{itemize}
\item An embedding layer transforms categorical inputs into 256-dimensional dense vectors.
\item Four stacked Bidirectional LSTM layers (512 units each), following prior character-level name-modeling work \citep{Xie2021, SoodLaohaprapanon2018}, capture temporal and contextual dependencies; we apply a dropout rate of 0.15 to mitigate overfitting.
\item Continuous geographic features are concatenated with LSTM output features for enhanced contextual understanding.
\item Final dense layer with softmax activation for multi-class classification.
\end{itemize}
\textbf{Training Details:}
\begin{itemize}
\item Optimizer: Adam ($\text{learning rate}=3.16\times 10^{-5}$)
\item Loss: Marginalized Cross-Entropy (below)
\item Fixed 10-epoch training schedule (no early stopping; per-epoch validation loss and White FPR trajectory converge by epoch 10), batch size of 512.
\end{itemize}

\paragraph{Marginalized Cross-Entropy Loss.}
STRATA is trained with a custom marginalized cross-entropy loss that reflects documented label ambiguity in the training sources rather than treating every label as equally certain. The output layer predicts over $K=8$ fine classes, $\mathcal{C}=\{$white, black, hisp, asian, nhpi, aian, other, multi$\}$, via a softmax $p_i \in \Delta^{K-1}$. Each record $i$ carries a source-provided provenance label $d_i$ (\texttt{race\_detail}) that determines an \emph{admissible set} $S(d_i)\subseteq\mathcal{C}$: exactly-measured labels map to a singleton (e.g.\ $S(\text{black})=\{$black$\}$), whereas categories a source is known to conflate map to the union of their members---FL/GA's combined Asian/Pacific-Islander field gives $S(\text{asian\_pi})=\{$asian, nhpi$\}$, and Georgia's ``Other'' checkbox gives $S(\text{other\_multi})=\{$other, multi$\}$. We encode $S(d_i)$ as a multi-hot mask $m_i\in\{0,1\}^K$ with $m_{i,k}=\mathbf{1}[k\in S(d_i)]$, and define the per-record loss as the negative log of the total probability mass the model assigns to the admissible set:
\begin{equation}
\mathcal{L}_i \;=\; -\log\!\Big(\textstyle\sum_{k=1}^{K} m_{i,k}\, p_{i,k}\Big)
\;=\; -\log\!\Big(\textstyle\sum_{k\in S(d_i)} p_{i,k}\Big).
\label{eq:marginalized_ce}
\end{equation}
When $S(d_i)$ is a singleton this reduces exactly to standard categorical cross-entropy; when it spans conflated members, the model is free to place mass on any admissible member without penalty, so it is never forced into spurious confidence on a distinction the source did not actually measure. This masking replaces the ad hoc $0.7$ sample-weight down-weighting applied to Georgia ``Other'' rows in an earlier model generation, giving the composite classes a principled training signal rather than a hand-tuned discount. For evaluation, fine probabilities are summed into the five-class scheme (nhpi folded into Asian; aian, other, and multi into Other) and scored against the coarse labels, keeping all reported metrics comparable across model generations. The practical consequence is measurable: NHPI recall is 3.8\% and AIAN recall is 3.4\%---these two fine-grained categories are recoverable only with additional, better-labeled training data, a limitation we return to in Section~5.

Hispanic origin is resolved from each source's ethnicity field during harmonization, \emph{before} coarse-class assignment: any record flagged Hispanic is labeled \texttt{hisp} regardless of its source race coding, so it never enters the Georgia ``Other'' admissible set. The residual ``Hispanic-origin leakage'' noted for Georgia refers only to records that Georgia records within its ``Other'' bucket \emph{without} an ethnicity flag; these are not separately identifiable and remain a small, acknowledged source of label noise in the \{other, multi\} mask rather than a systematic mislabeling. Table~\ref{tab:label_map} gives the full source-label to admissible-set to coarse-class mapping.

\begin{table}[htbp]
\centering
\caption{Provenance-label mapping: source race/ethnicity coding $\rightarrow$ fine admissible set $S(d)$ used in the marginalized loss $\rightarrow$ final coarse class.}
\label{tab:label_map}
\begin{tabular}{lll}
\toprule
Source label (\texttt{race\_detail}) & Admissible set $S(d)$ & Coarse class \\
\midrule
White / Black / Hispanic$^{\ast}$ & \{white\} / \{black\} / \{hisp\} & White / Black / Hispanic \\
FL, GA Asian/Pacific-Islander (\texttt{asian\_pi}) & \{asian, nhpi\} & Asian/PI \\
NC Asian (\texttt{asian}) & \{asian\} & Asian/PI \\
NHPI (NC ``P'') (\texttt{nhpi}) & \{nhpi\} & Asian/PI \\
AIAN (FL/NC code 1; GA native) (\texttt{aian}) & \{aian\} & Other \\
GA ``Other'' checkbox (\texttt{other\_multi}) & \{other, multi\} & Other \\
FL/NC multiracial (\texttt{multi}) & \{multi\} & Other \\
Other (\texttt{other}) & \{other\} & Other \\
\bottomrule
\end{tabular}
\parbox{\textwidth}{\small \textit{Note:} $^{\ast}$Hispanic is assigned from the ethnicity field prior to coarse-class construction and takes precedence over source race coding. Composite admissible sets ($|S(d)|>1$) arise only where a source genuinely conflates categories; all other labels are singletons and reduce the loss to standard cross-entropy.}
\end{table}

\subsection{STRATA Ensemble (XGBoost Post-Filtering)}

To further enhance performance, the STRATA ensemble uses XGBoost as a secondary model to post-process STRATA base predictions. This step captures complex nonlinear interactions and systematically corrects residual errors from the neural predictions \citep{Argyle2023, ZestAIRelease2022}.

\paragraph{Rationale.}
\begin{itemize}
\item \textbf{Complementary Strengths.} While LSTM models excel at sequential pattern modeling \citep{Xie2021}, XGBoost captures intricate nonlinear patterns in tabular data \citep{Argyle2023, ZestAIRelease2022}.
\item \textbf{Interpretability.} XGBoost provides explicit feature importances, indicating which neural predictions and demographic features drive improved predictions.
\item \textbf{Robust to Imperfections.} Gradient boosted trees effectively handle noisy predictions, deemphasizing irrelevant features.
\end{itemize}

\paragraph{Implementation Details.} The post-filter is an XGBoost gradient-boosted-tree classifier (\texttt{xgboost} 2.x, \texttt{tree\_method="hist"}) trained on the concatenation of the eight fine-class probabilities from STRATA base and the same block-group/tract racial-composition and income-decile features the base model consumes, with the coarse five-class label as target (\texttt{objective="multi:softprob"}, \texttt{num\_class=5}). Rather than a per-run hyperparameter search---which would not be reproducible from released artifacts---we fix a single configuration: \texttt{n\_estimators}$=400$, \texttt{max\_depth}$=6$, \texttt{learning\_rate}$=0.1$, \texttt{subsample}$=0.8$, \texttt{colsample\_bytree}$=0.8$, and XGBoost's default $L_1$/$L_2$ regularization ($\texttt{reg\_alpha}=0$, $\texttt{reg\_lambda}=1$). The post-filter is trained with 5-fold stratified cross-validation (\texttt{StratifiedKFold}, \texttt{shuffle=True}, \texttt{random\_state=42}); the five fold models are retained and their softmax outputs averaged at inference, so the deployed ensemble is the mean of five independently fit boosters. All settings are fixed constants in the released training code, making the ensemble stage reproducible from the archived base-model probabilities and locked software environment.

\emph{Stacking procedure (out-of-fold).} The post-filter is trained on \emph{out-of-fold} base-model probabilities so that it never sees in-sample (over-confident) base predictions. The training rows are partitioned into five stratified folds; for each fold, a fold-specific LSTM is trained on the other four folds and used only to generate base probabilities for the held-out fold, yielding out-of-fold probabilities for every training row. The stacker is then fit on those out-of-fold probabilities using the same stratified five-fold partition (\texttt{StratifiedKFold}, seed 42): for each fold $j$ an XGBoost booster is trained on the other four folds' out-of-fold probabilities, yielding five boosters. At inference, each of the five boosters receives base probabilities from the single full-data BiLSTM (trained on all rows, unchanged) and their predicted class-probability vectors are averaged. All principal voter- and PPP-holdout ensemble results (Tables~\ref{tab:holdout2_overall_performance}--\ref{tab:confusion}, \ref{tab:geo_ablation}, \ref{tab:ppp-national-results}, \ref{tab:aggregate}) derive from the resulting frozen prediction set (checksum in the artifact release); the historical geographic-transfer experiments of Table~\ref{tab:loso} use their separately frozen model generation (Table~\ref{tab:cohort_flow}) with an in-sample stacker, so the out-of-fold correction does not apply to them and their absolute values are not directly comparable. The cross-fitting uses row-level stratified folds, which removes in-sample leakage \emph{at the row level}; we do not group folds by normalized name--address entity, so duplicate or near-duplicate records (2.98\% of training rows share a name--tract key---2.2\% in the voter block but 19.8\% within the smaller PPP block) may still appear across folds. We disclose this rather than claim protection against it; entity-grouped cross-fitting is a natural refinement, and we do not describe the folds as fully leakage-proof for related PPP entities. This out-of-fold design replaces an in-sample-stacker configuration used in earlier versions of this manuscript; the correction changes the reported ensemble metrics only marginally (e.g., voter White FPR $0.175 \rightarrow 0.178$, accuracy $0.887 \rightarrow 0.888$).

\subsection{Summary of Proposed Methods}

The integrated STRATA framework provides enhanced accuracy by jointly modeling linguistic, geographic, and socioeconomic interactions. This combined approach significantly outperforms existing Bayesian techniques (BISG, BIFSG) and standalone neural-network methods \citep{Elliott2009, Voicu2018, SoodLaohaprapanon2018}, delivering both higher accuracy and reduced classification bias.

\subsection{Model Input Comparison}

Race Probabilities (RPs) defined as a set of probabilities for an individual to be of one of 5 race/ethnicity options: White, African American, Hispanic, Asian, and Other.

\begin{table}[htbp]
\centering
\caption{Illustrative Data/Feature Requirements of Various Models (from Literature)}
\label{tab:data_features}
\begin{tabular}{||p{0.22\linewidth}|p{0.15\linewidth}|p{0.63\linewidth}||}
\toprule
Method & Input feature count & Main feature list \\
\midrule
BISG & 10 & Last name implied RPs, Tract implied RPs \\
BIFSG & 15 & First name implied RPs, Last name implied RPs, Tract implied RPs \\
Argyle \& Barber RF & 23 & BISG posteriors (5), income, education, homeownership, home value, tract composition, plus voter-file-only fields: party registration, vote propensity, campaign donations, voter-file age, $\cdots$ \\
ZRP (Zest) & $\sim$203 & First, Last, Middle names, $\sim$200 engineered tract/ACS variables \\
STRATA base & name + 13 & First, middle, last name (character-level); block-group and tract racial composition (5 rates each); block-group and tract income deciles; block-group population \\
STRATA ensemble & name + 13 & Same inputs as STRATA base (XGBoost consumes the base model's class probabilities plus the 13 geo features) \\
\bottomrule
\end{tabular}
\parbox{\textwidth}{\small \textit{Note: Argyle \& Barber and ZRP feature lists were truncated; full list available at source.}}
\end{table}

\section{Results and Comparisons}

We compare STRATA base and STRATA ensemble against baseline LSTM, BISG (surgeo), BIFSG (surgeo), ZRP, and a faithful replication of Argyle \& Barber's BISG+random-forest correction, all applied to the same initial matched voter holdout ($N=981{,}288$), with BISG and BIFSG metrics computed on their respective scored subsets.

\subsection{Overall Performance Metrics}

Table \ref{tab:holdout2_overall_performance} summarizes the overall performance metrics for the matched voter holdout.

\begin{table}[htbp]
\centering
\caption{Overall Performance Metrics on the Matched Voter Holdout ($N=981{,}288$)}
\label{tab:holdout2_overall_performance}
\begin{tabular}{lccccc}
\toprule
Model & Accuracy & F1-Score & Precision & Macro OvR FPR & Coverage \\
\midrule
LSTM (name only) & 0.858 & 0.845 & 0.848 & 0.0612 & 100\% \\
STRATA base & 0.885 & 0.875 & 0.875 & 0.0487 & 100\% \\
ZRP & 0.863 & 0.851 & 0.854 & 0.0587 & 100\% \\
BIFSG (Surgeo)$^{\dagger}$ & 0.859 & 0.846 & 0.847 & 0.0746 & 77.5\% \\
BISG (Surgeo)$^{\dagger}$ & 0.823 & 0.812 & 0.815 & 0.0755 & 89.6\% \\
A\&B (BISG+RF)$^{\ddagger}$ & 0.822 & 0.814 & 0.814 & 0.0687 & 100\% \\
\textbf{STRATA ensemble} & \textbf{0.888} & \textbf{0.877} & \textbf{0.879} & \textbf{0.0475} & \textbf{100\%} \\
\bottomrule
\end{tabular}
\parbox{\textwidth}{\small \textit{Note:} F1-Score and Precision are weighted averages. ``Macro OvR FPR'' is the macro-average of the five per-class one-vs-rest FPRs; we report it as a secondary descriptive statistic only, because a class can earn a low one-vs-rest FPR simply by being predicted rarely (the ``Other'' class, recall 0.056, is an example). White FPR, macro-F1, balanced accuracy, and the aggregate share error of Section~\ref{sec:aggregate} are the more substantively meaningful summaries. All methods are applied to the same initial matched holdout; BISG and BIFSG metrics are computed on their respective scored subsets. $^{\dagger}$Under the specified Surgeo implementation and reference lists, BISG and BIFSG return a prediction for 89.6\% and 77.5\% of the cohort, respectively (surname/first-name not in the reference list yields no output); their accuracy is computed over those scored subsets. Treating missing output as an error on the full cohort, their accuracies fall to approximately 0.73 and 0.66. $^{\ddagger}$A\&B is a faithful replication of \citet{Argyle2023}; its 23 predictors include voter-file-only fields (party registration, vote propensity, campaign donations, voter-file age) that do not exist in lending or housing records, so it is included for completeness but is not directly transferable to those domains without alternative covariates (Section~2.6). STRATA ensemble attains the highest accuracy and lowest macro OvR FPR at 100\% conditional model-output coverage among eligible tract-resolved records; its macro-F1 is 0.689 and balanced accuracy (mean per-class recall) 0.679, reflecting the weak recovery of the heterogeneous ``Other'' residual (Table~\ref{tab:per_class}).}
\end{table}

STRATA base (Acc: 0.885) outperforms standalone LSTM (0.858) and ZRP (0.863) at 100\% conditional model-output coverage.\footnote{Reporting convention: model performance metrics (accuracy, false positive rates, precision, recall) are given as decimals in the body and tables and as percentages in the abstract, introduction, and conclusion; group differences (disparities and aggregate share errors) are given in percentage points (pp).} The STRATA ensemble achieves the highest accuracy (0.888) and the lowest macro OvR FPR (0.0475) among all methods evaluated at 100\% conditional model-output coverage among eligible tract-resolved records. The coverage column is central to the fair-lending interpretation: under the specified Surgeo implementation, BISG and BIFSG return metrics only on the 77--90\% of records for which their surname/first-name reference lists yield a prediction; the unscored individuals are absent from any disparity analysis built on the proxy unless a fallback is imposed. STRATA assigns a prediction to every record.

\subsection{False Positive Rate and Bias Reduction}

Table \ref{tab:holdout2_fpr_race} presents the recalculated FPR for each racial group on the matched voter holdout.

\begin{table}[htbp]
\centering
\caption{False Positive Rate by Race/Ethnicity on the Matched Voter Holdout ($N=981{,}288$)}
\label{tab:holdout2_fpr_race}
\begin{tabular}{lcccccc}
\toprule
Model & White & Black & Hispanic & Asian & Other & Coverage \\
\midrule
LSTM (name only) & 0.235 & 0.042 & 0.023 & 0.006 & 0.001 & 100\% \\
STRATA base & 0.184 & 0.032 & 0.021 & 0.005 & 0.002 & 100\% \\
ZRP & 0.225 & 0.029 & 0.027 & 0.011 & 0.001 & 100\% \\
BIFSG (Surgeo) & 0.312 & 0.036 & 0.022 & 0.002 & 0.001 & 77.5\% \\
BISG (Surgeo) & 0.280 & 0.063 & 0.028 & 0.005 & 0.001 & 89.6\% \\
A\&B (BISG+RF) & 0.227 & 0.067 & 0.040 & 0.008 & 0.002 & 100\% \\
\textbf{STRATA ensemble} & \textbf{0.178} & \textbf{0.032} & \textbf{0.021} & \textbf{0.005} & \textbf{0.001} & \textbf{100\%} \\
\bottomrule
\end{tabular}
\parbox{\textwidth}{\small \textit{Note:} All methods were applied to the same initial holdout; BISG/BIFSG FPRs use their scored subsets. Each row's five per-class one-vs-rest FPRs are the entries macro-averaged in the ``Macro OvR FPR'' column of Table~\ref{tab:holdout2_overall_performance}. White FPR is the rate at which non-White individuals are misclassified as White. Coverage denotes the fraction of records the method scores; BISG/BIFSG values are computed only over scored records. A\&B requires 23 voter-file-only predictors and is included for completeness, not as a deployable comparator (Section~2.6).}
\end{table}

\begin{figure}[tbp]
\centering
\includegraphics[width=0.9\textwidth]{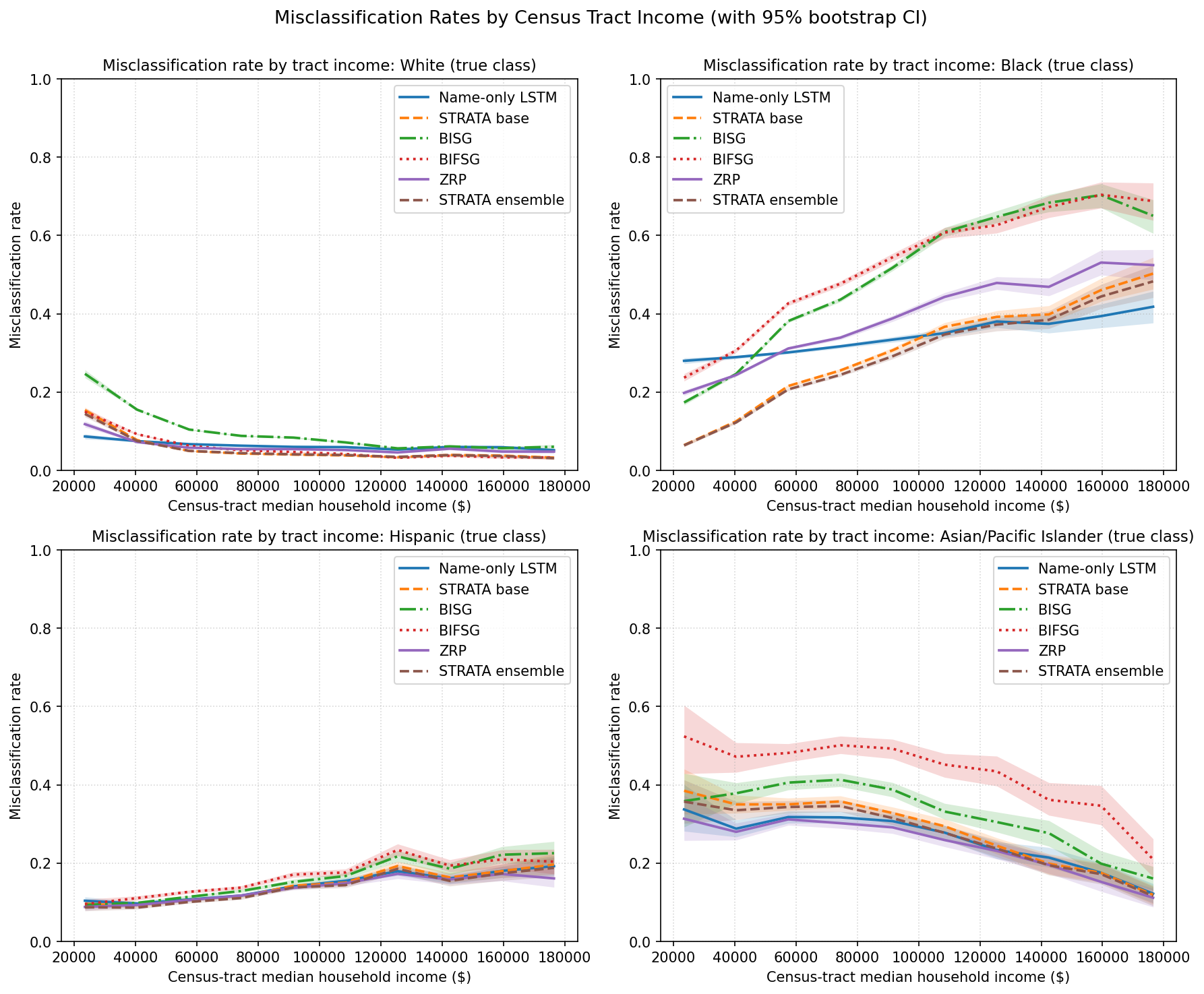}
\caption{Per-class misclassification rate as a function of census-tract median income on the matched voter holdout ($N=981{,}288$), for each model. Curves plot $P(\widehat{Y}\neq c \mid Y=c)$ within fixed tract-income bins; shaded regions are 95\% bootstrap confidence intervals (1{,}000 resamples per bin). BISG and BIFSG display stronger income-associated variation, most pronounced for Black records (and non-monotonic for Hispanic and, at the highest incomes, declining for Asian), whereas STRATA (ensemble) and STRATA base remain lower across much of the income range.}
\label{fig:misclass1}
\end{figure}

The STRATA ensemble attains the lowest White FPR (0.178) of any evaluated method, below ZRP (0.225), A\&B (0.227), BISG (0.280), and BIFSG (0.312). Because the principal application risk in this setting is the absorption of non-White records into the White predicted class, we report the White one-vs-rest FPR as a \emph{targeted} error measure. We do not interpret the one-vs-rest FPRs of the minority predicted classes as errors experienced by members of those classes: $\mathrm{FPR}_{\text{Black}} = P(\widehat{Y}=\text{Black}\mid Y\neq\text{Black})$ counts false Black predictions among non-Black records, not errors incurred by Black individuals. Errors experienced by each class are assessed separately through class-specific recall, false-negative rate, precision, F1, and probabilistic accuracy (reported through classwise Brier scores), in Table~\ref{tab:per_class} and the confusion matrix of Table~\ref{tab:confusion}. The Brier score is a proper probabilistic score reflecting both calibration and refinement; a dedicated reliability diagnostic is provided in the artifact release. The nearest competitor on White FPR, ZRP (0.225), requires $\sim$203 engineered ACS features, and the next, Argyle \& Barber's correction (0.227), uses voter-file-specific covariates that are ordinarily unavailable in mortgage, deed, and small-business-loan records, limiting direct deployment in those domains. Under the specified Surgeo implementation, BISG and BIFSG return predictions for only 89.6\% and 77.5\% of the cohort respectively (their per-class values are computed over those scored subsets), whereas STRATA scores every record. Figure~\ref{fig:misclass1} shows how misclassification rates vary with census-tract income across racial groups, illustrating that the SES-correlated error pattern affecting BISG is substantially attenuated under STRATA.

To characterize the errors \emph{experienced} by each group (as opposed to the White-absorption risk captured by the one-vs-rest FPR), Table~\ref{tab:per_class} reports the full per-class performance of the STRATA ensemble on the matched voter holdout, and Table~\ref{tab:confusion} gives the underlying $5\times5$ confusion matrix. Two patterns warrant emphasis. First, recall is high and precision strong for White (0.952/0.910), Hispanic (0.884/0.859), and Black (0.797/0.839), but the ``Other'' class---an administrative residual aggregating AIAN (5{,}094 records), source-``Other'' (14{,}352), and Georgia's combined other/multi checkbox (6{,}684), with pure multiracial records excluded from the evaluation set (Section~\ref{sec:taxonomy})---has recall of only 0.056: the model overwhelmingly reassigns these records to White (13{,}040 of 26{,}130) or Black (6{,}065). This poor recall is consistent with the heterogeneity and limited learnability of the residual ``Other'' category, compounded by its low prevalence and label-source inconsistencies, and we report it transparently rather than suppress it. Second, the White false-negative rate (0.048) is substantially lower than the minority-class false-negative rates, showing that errors are concentrated in the failure to recover minority labels; the confusion matrix further shows that a large share of those failures are assignments into the White predicted class. We therefore treat the White one-vs-rest FPR (0.178) as the most decision-relevant single measure of majority-class absorption for this application, but not as a sole summary of fairness: we report it alongside per-class recall, F1, and Brier, and the macro-F1 (0.689) and balanced accuracy (0.679), which are sensitive to the poor recovery of the ``Other'' residual that a low ``Other'' one-vs-rest FPR would otherwise mask.

\begin{table}[htbp]
\centering
\caption{Per-Class Performance of the STRATA Ensemble on the Matched Voter Holdout ($N=981{,}288$). FPR is one-vs-rest. Brackets give 1000-iteration bootstrap 95\% CIs.}
\label{tab:per_class}
\begin{tabular}{lccccccc}
\toprule
Class & Prevalence & Precision & Recall & FNR & F1 & FPR & Brier \\
\midrule
White & 0.655 & 0.910 [.909,.911] & 0.952 [.951,.952] & 0.048 & 0.930 [.930,.931] & 0.178 & 0.071 \\
Black & 0.173 & 0.839 [.837,.841] & 0.797 [.795,.799] & 0.203 & 0.817 [.816,.819] & 0.032 & 0.046 \\
Hispanic & 0.126 & 0.859 [.857,.860] & 0.884 [.882,.886] & 0.116 & 0.871 [.870,.872] & 0.021 & 0.026 \\
Asian & 0.020 & 0.738 [.732,.744] & 0.706 [.700,.713] & 0.294 & 0.722 [.717,.727] & 0.005 & 0.008 \\
Other & 0.027 & 0.585 [.566,.604] & 0.056 [.053,.059] & 0.944 & 0.103 [.098,.107] & 0.001 & 0.024 \\
\bottomrule
\end{tabular}
\parbox{\textwidth}{\small \textit{Note:} Prevalence is the empirical class share in the holdout. Recall/FNR measure errors experienced by each true class; precision measures reliability of each predicted label; FPR (one-vs-rest) measures absorption of other classes into the predicted label; Brier is the per-class one-vs-rest Brier score (mean squared error of the predicted class probability), a proper probabilistic score reflecting both calibration and refinement. Macro-averaged over the five classes, the ensemble attains macro-F1 $0.689$ and balanced accuracy (mean per-class recall) $0.679$. The ``Other'' row's low recall reflects its status as a heterogeneous administrative residual, not a coherent population (Section~\ref{sec:taxonomy}).}
\end{table}

\begin{table}[htbp]
\centering
\caption{Confusion Matrix for the STRATA Ensemble on the Matched Voter Holdout ($N=981{,}288$). Rows are true class; columns are predicted class.}
\label{tab:confusion}
\begin{tabular}{lccccc|c}
\toprule
True $\backslash$ Pred & White & Black & Hispanic & Asian & Other & Total \\
\midrule
White & 611{,}240 & 16{,}803 & 12{,}783 & 1{,}188 & 335 & 642{,}349 \\
Black & 32{,}213 & 135{,}178 & 1{,}695 & 330 & 239 & 169{,}655 \\
Hispanic & 11{,}673 & 2{,}279 & 109{,}411 & 371 & 53 & 123{,}787 \\
Asian & 3{,}516 & 820 & 949 & 13{,}668 & 414 & 19{,}367 \\
Other & 13{,}040 & 6{,}065 & 2{,}592 & 2{,}965 & 1{,}468 & 26{,}130 \\
\midrule
Total & 671{,}682 & 161{,}145 & 127{,}430 & 18{,}522 & 2{,}509 & 981{,}288 \\
\bottomrule
\end{tabular}
\parbox{\textwidth}{\small \textit{Note:} Diagonal entries are correct predictions. The dominant off-diagonal mass is misassignment of true-minority records into the predicted-White column (col.\ 1), which the White one-vs-rest FPR (0.178) summarizes.}
\end{table}

\subsection{Geolocation Ablation}

To isolate the contribution of the geolocation features, we compare STRATA base against an otherwise-identical name-only configuration (same architecture, hyperparameters, holdout, and random seed, with tract and income features removed). Table~\ref{tab:geo_ablation} reports the result.

\begin{table}[htbp]
\centering
\caption{Geolocation Ablation on the Matched Voter Holdout}
\label{tab:geo_ablation}
\begin{tabular}{lccc}
\toprule
Configuration & Accuracy & Macro F1 & White FPR \\
\midrule
Name-only LSTM & 0.8577 & 0.653 & 0.235 \\
STRATA base (name + geo + income) & 0.8854 & 0.687 & 0.184 \\
\bottomrule
\end{tabular}
\parbox{\textwidth}{\small \textit{Note:} Both configurations use identical architecture, hyperparameters, holdout split, and seed; only the geolocation and income features differ.}
\end{table}

Adding geolocation and income features contributes $+2.8$ percentage points of accuracy and reduces the White FPR by $5.1$ points (0.235 to 0.184). This directly isolates the effect of the tract-referenced feature integration that distinguishes STRATA from name-only sequence models: the majority-class absorption that drives the shrinkage characterized by \citet{Xin2026} is measurably reduced once the model can condition on the individual's own residential context.

\subsection{Sensitivity to Training Geography}
\label{sec:loso}

A natural concern is that STRATA's performance could depend on geographic overlap between its training states and its evaluation set, rather than on a genuinely transferable name--geography relationship. We are explicit about what the principal model does and does not test. Because the principal STRATA model includes national PPP observations during training (Section~3.1), its held-out PPP result (Table~\ref{tab:ppp-national-results}) is an \emph{in-source} test---the held-out records are disjoint by record but share the PPP source distribution---not a strict unseen-geography test. To assess geographic transfer independently of PPP training, we use two auxiliary models trained with \emph{no} PPP observations at all, M1 (FL+NC voter only) and M2 (FL+GA+NC voter only), and evaluate them on the national PPP cohort; for these models every PPP record is genuinely out-of-source and predominantly out-of-training-geography. Table~\ref{tab:loso} reports the result, together with a Georgia training-state ablation.

The Georgia contrast (M1 vs.\ M2, and M3 vs.\ M4) is a \emph{training-state ablation}: it adds or removes the Georgia voter population from training while holding architecture, features, split rule, and epoch schedule fixed. It is not a leave-one-state-out validation, because the evaluation sets are national PPP (M1/M2) and New York PPP (M3/M4), not a held-out Georgia cohort. For the New York comparison (M3/M4), New York PPP records were held out entirely---no New York record entered base-model fitting, stacker fitting, model selection, or the preprocessing statistics (income-decile bins)---so ``+PPP'' for M3/M4 means the national PPP file \emph{excluding} New York; the New York PPP set is therefore a genuine unseen-state evaluation.

\begin{table}[htbp]
\centering
\caption{Historical auxiliary sensitivity analysis: training-geography ablation. Ensemble accuracy and White FPR. M1/M2 are scored on the national PPP file ($n=910{,}762$, predominantly states unseen in training); M3/M4 are scored on a fully held-out New York PPP set ($n=35{,}974$, a state never in training). These experiments use an earlier data and stacker generation and are interpreted only as within-generation sensitivity comparisons.}
\label{tab:loso}
\begin{tabular}{llccc}
\toprule
Model & Train states (PPP) & Score set & Accuracy & White FPR \\
\midrule
M1 & FL+NC (no PPP) & National PPP & 0.850 & 0.138 \\
M2 & FL+GA+NC (no PPP) & National PPP & 0.859 & 0.128 \\
\midrule
M3 & FL+NC (+PPP) & New York PPP (OOS) & 0.871 & 0.073 \\
M4 & FL+GA+NC (+PPP) & New York PPP (OOS) & 0.876 & 0.069 \\
\bottomrule
\end{tabular}
\parbox{\textwidth}{\small \textit{Note:} Ensemble (LSTM$+$XGBoost) results; all four models use the identical architecture and hyperparameters of Section~3. ``(+PPP)''/``(no PPP)'' indicates whether the national PPP file was included in training. M1 vs.\ M2 and M3 vs.\ M4 each isolate the marginal contribution of the Georgia voter population, holding everything else fixed.}
\end{table}

Two points follow. Removing Georgia entirely (M1, M3) costs only $0.5$--$0.9$ percentage points of accuracy and $0.4$--$1.0$ points of White FPR relative to the full three-state model (M2, M4): the limited change between M1 and M2 (and between M3 and M4) indicates that the model's transfer performance is not critically dependent on the inclusion of Georgia voter records. The models retain $85$--$87\%$ accuracy scoring on populations drawn overwhelmingly (M1/M2, national PPP) or entirely (M3/M4, New York) from states outside their training geography. The New York comparison (M3/M4) is a strict unseen-state test---no New York record appears anywhere in training---and STRATA still attains $87\%$ accuracy with a White FPR below $0.075$. These experiments provide evidence of geographic transfer beyond the three voter-training states; they do not, of course, isolate the marginal contribution of Florida or North Carolina, only of Georgia.

\subsection{Statistical Significance}

We assess significance with paired tests conducted on each pair's common support. McNemar's test was completed for the STRATA ensemble versus BISG, BIFSG, and A\&B, each on the records both methods score, and rejects the null of equal error rates in every case at $p < 10^{-300}$ (vs.\ BISG $\chi^2 = 32{,}963$, $n=879{,}461$; vs.\ BIFSG $\chi^2 = 12{,}098$, $n=760{,}827$; vs.\ A\&B $\chi^2 = 36{,}136$, $n=981{,}288$). ZRP was compared with the STRATA ensemble using a paired bootstrap (1{,}000 resamples) of the White-FPR difference, which yields a 95\% confidence interval of $[4.60, 4.82]$ percentage points in STRATA's favor ($p < 0.001$). Ordinary (record-level) bootstrap resampling of the voter holdout yields an ensemble accuracy of 0.8876 (95\% CI $[0.8870, 0.8882]$) and a White FPR of 0.1783 (95\% CI $[0.1771, 0.1795]$). Because records sharing a Census tract are not independent, we also report a bootstrap clustered by tract ($11{,}320$ tracts resampled as units), which appropriately widens the intervals: accuracy 0.8876 (95\% CI $[0.8864, 0.8887]$) and White FPR 0.1784 (95\% CI $[0.1747, 0.1822]$). The differences remain highly precise even under clustering; the practical significance rests on the magnitude and coverage arguments above rather than on statistical significance alone.

\subsection{Efficacy of STRATA Base}

The standalone STRATA base model offers unique value despite the STRATA ensemble achieving the highest accuracy. Its deep learning architecture learns complex interactions between name components and geographic context, unlike Bayesian methods' reliance on pre-calculated probabilities and independence assumptions. Concatenating the block-group and tract covariates with the final BiLSTM representation before the softmax (Section~3.2), so that a single discriminative objective learns the name--geography interaction, is a principal architectural contribution of this paper relative to Bayesian proxies that combine the two through a fixed factorization. Achieving 0.885 accuracy and a substantial reduction in White-class absorption as a single model makes STRATA base a strong option where model simplicity is valued.

\subsection{Parsimony and Deployability}

The best results came from using STRATA base predictions as input to the XGBoost model. Boosted decision trees were also present in the next best performing models: Argyle \& Barber's BISG into random forest \citep{Argyle2023} and ZRP \citep{ZestGitHubZRP}. The decisive difference is the input requirement (Table~\ref{tab:data_features}): STRATA uses a name plus 13 public Census features, available for any US address from public data. ZRP requires $\sim$203 engineered ACS features, a substantially heavier feature pipeline to reproduce. Argyle \& Barber's correction requires 23 predictors, roughly half of which---party registration, modeled vote propensity, campaign-donation history, and voter-file age---exist only in voter files and cannot be supplied in mortgage, deed, or small-business-loan records. BIFSG additionally requires a first name beyond BISG's surname-and-geography inputs, and both return no output for records outside their reference lists. STRATA therefore achieves competitive or superior performance from the smallest and most broadly available \emph{input} set of any evaluated method---a name plus 13 public Census features available in the domains where aggregate proxy analysis is legally and methodologically appropriate (lending, housing, loan-level compliance). We describe STRATA as parsimonious in required inputs, not in learned parameters: the BiLSTM component has $66{,}156{,}626$ trainable weights ($\approx$66.2M; Section~3.2). The advantage is deployability and reproducibility of inputs, not model size.

\subsection{PPP-Based National Validation}

To validate model performance nationally, we conducted a PPP-based evaluation using the Paycheck Protection Program dataset, a federal loan initiative under the CARES Act. The publicly available PPP dataset includes borrower names, locations, and self-identified race and ethnicity for millions of applicants across all 50 U.S. states and Washington, D.C., providing a rare, large-scale national benchmark.

\paragraph{Person-name cohort restriction.} A PPP \texttt{BorrowerName} may denote an individual or a business entity (LLC, nonprofit, trade name), and only for individuals does the self-reported demographic label plausibly describe the person the name encodes. We therefore restrict the national validation to a person-name cohort via a two-stage filter: (Stage 1) \texttt{BusinessType} $\in$ \{Sole Proprietorship, Self-Employed Individuals, Independent Contractors, Single Member LLC\}; (Stage 2) a person-name heuristic requiring at least two alphabetic tokens and no business keyword. The token pattern \texttt{[A-Za-z'-]+} preserves hyphenated and apostrophe surnames as single tokens (e.g.\ ``Abdel-Hadi,'' ``O'Neill''), and---because STRATA encodes a dedicated middle-name field---the filter retains individuals whose borrower name includes a middle name or a second surname rather than discarding them. Of the 125{,}081 tract-resolved PPP holdout records, 124{,}914 (99.9\%) pass both stages; only 167 (0.1\%) are removed (162 containing a business keyword and 5 with fewer than two alphabetic tokens). All reported PPP metrics below are computed on this $N=124{,}914$ person-name cohort, which comprises Sole Proprietorship (109{,}813; 87.9\%) and Self-Employed Individuals (15{,}101; 12.1\%) records only; no Single Member LLC or Independent Contractor borrower string survives Stage 2, as each either carries an entity keyword or resolves to a business name. This is a name-and-address robustness evaluation on self-identified small-business borrowers; a standardized street address is not by itself evidence of residential occupancy, and residual entity contamination cannot be fully excluded. The person-name restriction is rule-based and was not independently validated through manual annotation; residual person/entity classification error is therefore possible. We rely on the structured \texttt{BusinessType} field, rather than a stricter name-token rule, for person/entity separation: an exactly-two-token requirement would additionally exclude 13{,}852 borrowers who report a middle name or two surnames, and its exclusion rate is markedly higher among Hispanic and Asian cohort members (Table~\ref{tab:excl2tok}), introducing a race-correlated selection into the cohort without reducing entity contamination, since all 13{,}852 self-report an individual \texttt{BusinessType}.

\begin{table}[htbp]
\centering
\caption{Race-correlated selection avoided by the person-name filter. Within the final PPP person-name cohort ($N=124{,}914$), the share of each self-reported class that an \emph{exactly}-two-token rule would exclude---i.e., records whose borrower name has a middle name or a second surname. Every excluded record self-reports an individual \texttt{BusinessType}; the more permissive at-least-two-token rule retains them.}
\label{tab:excl2tok}
\begin{tabular}{lrrr}
\toprule
Class & Cohort $N$ & Excluded (exact 2-token) & Exclusion rate \\
\midrule
White & 47{,}115 & 7{,}301 & 15.5\% \\
Black & 61{,}067 & 3{,}226 & 5.3\% \\
Hispanic & 9{,}016 & 1{,}846 & 20.5\% \\
Asian & 5{,}919 & 1{,}235 & 20.9\% \\
Other & 1{,}797 & 244 & 13.6\% \\
\midrule
Overall & 124{,}914 & 13{,}852 & 11.1\% \\
\bottomrule
\end{tabular}
\end{table}

\begin{table}[htbp]
\centering
\caption{Performance on the National PPP Person-Name Holdout ($N=124{,}914$)}
\label{tab:ppp-national-results}
\begin{tabular}{lccccc}
\toprule
\textbf{Model} & \textbf{Accuracy} & \textbf{F1-Score} & \textbf{Precision} & \textbf{Recall} & \textbf{Macro OvR FPR} \\
\midrule
LSTM (Name-Only) & 0.8089 & 0.8034 & 0.8003 & 0.8089 & 0.0646 \\
STRATA base & 0.8760 & 0.8707 & 0.8700 & 0.8760 & 0.0405 \\
\textbf{STRATA ensemble} & \textbf{0.8845} & \textbf{0.8785} & \textbf{0.8789} & \textbf{0.8845} & \textbf{0.0379} \\
ZRP (XGBoost Ensemble) & 0.8073 & 0.8040 & 0.8200 & 0.8073 & 0.0590 \\
\bottomrule
\end{tabular}
\vspace{0.1cm}
\small{\textit{Note: F1-Score and Precision are weighted averages; Recall equals Accuracy in this single-label multiclass setting; the Macro OvR FPR is the macro-average of the five per-class one-vs-rest FPRs, matching the convention used in Tables~\ref{tab:holdout2_overall_performance}--\ref{tab:holdout2_fpr_race}. All four methods score every record in this cohort (100\% coverage); ZRP was re-scored per-record to report weighted F1/Precision.}}
\end{table}

These results are a geographically national, in-source held-out test, and should not be conflated with the voter-holdout results of Tables~\ref{tab:holdout2_overall_performance}--\ref{tab:holdout2_fpr_race}. Because PPP records also appear in STRATA's training corpus (Section~3.1), the held-out PPP split shares the source distribution of part of the training data; it is therefore not a cross-domain test, but it does span all 50 states and a very different population (self-identifying small-business borrowers rather than registered voters), so it measures how the model generalizes geographically and to a distinct applicant population within a source it has partly seen. On this national person-name cohort the STRATA ensemble attains the highest performance at 0.885 accuracy with a White FPR of 0.0981 and the lowest macro OvR FPR (0.0379), closely followed by STRATA base (0.0405 overall FPR). Both substantially reduce misclassification errors compared to the name-only LSTM (overall FPR 0.0646). This national held-out result provides the most geographically broad deployment-oriented evaluation in the paper, while remaining an in-source and selected-cohort test (PPP records also appear in training, and the cohort is restricted to race-disclosing, successfully geocoded, person-name small-business borrowers).

\begin{figure}[tbp]
\centering
\includegraphics[width=0.9\textwidth]{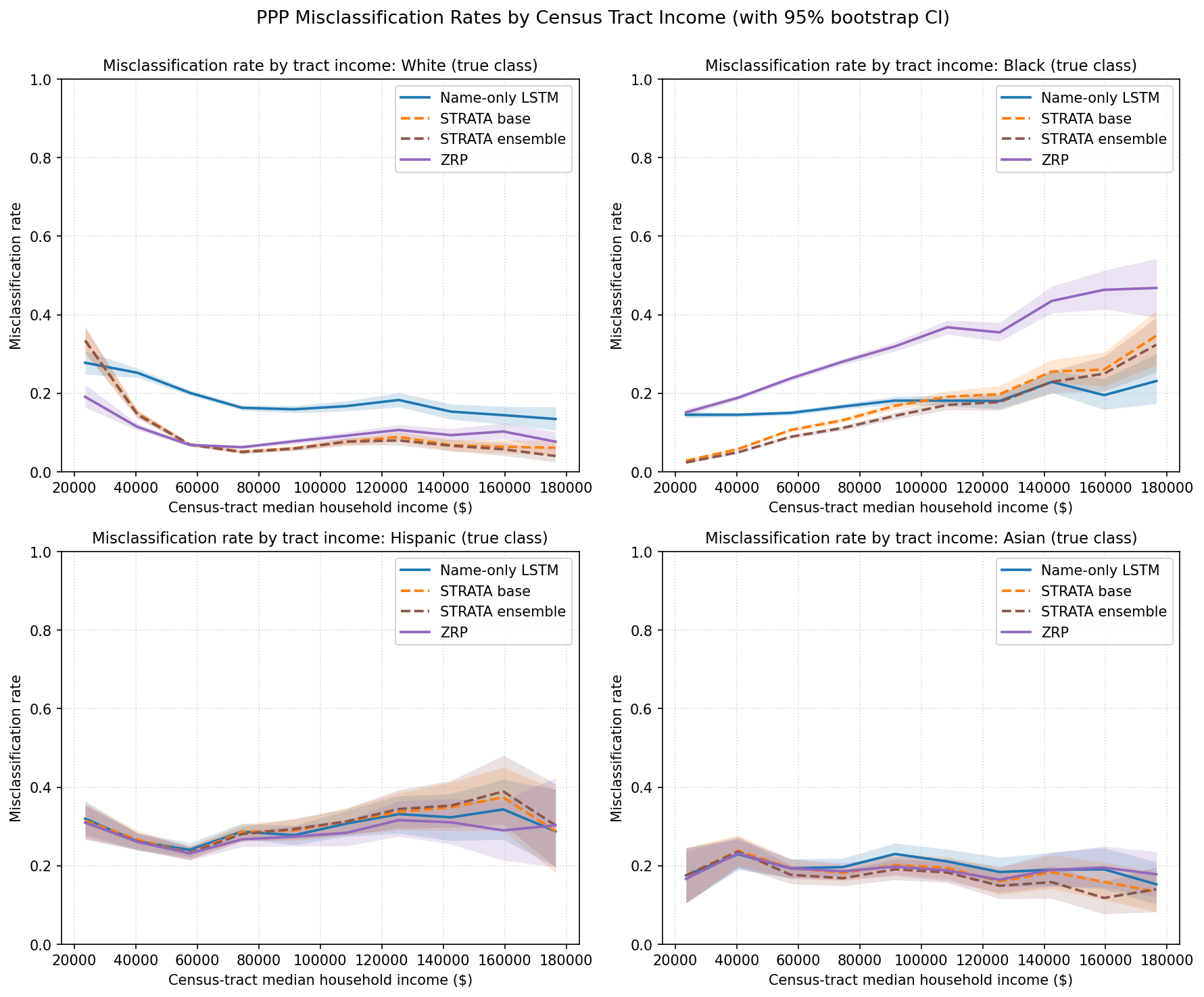}
\caption{Per-class misclassification rate as a function of census-tract median income on the national PPP person-name cohort ($N=124{,}914$; Section~4.8), for the name-only LSTM, STRATA base, the STRATA ensemble, and ZRP. Curves plot $P(\widehat{Y}\neq c \mid Y=c)$ within fixed tract-income bins; shaded regions are 95\% bootstrap confidence intervals (1{,}000 resamples per bin). ZRP exhibits a pronounced positive income gradient in Black-class misclassification within this cohort, whereas the STRATA configurations remain lower and flatter.}
\label{fig:misclass2}
\end{figure}

Figure~\ref{fig:misclass2} compares misclassification rates by income level and race across these models on the PPP data. The cohort spans all 50 states and the District of Columbia (51 jurisdictions). Accuracy varies across states, reflecting differences in demographic composition and name distribution: among states with at least 500 cohort records it is highest in lower-diversity states (e.g., Iowa, Nebraska) and lowest in more diverse states (e.g., Washington, Nevada, Arizona). To avoid over-reading small cohorts, we restrict state-level ranking claims to states with $N\geq500$ (37 states); per-state accuracy and balanced accuracy with sample sizes, for all states meeting this threshold, are reported in Appendix~\ref{app:perstate}. Very small state cohorts (e.g., Hawaii, Vermont, Maine, Alaska, each $N<300$) are excluded from these rankings, though Hawaii remains informative for the five-class-collapse limitation discussed in Section~5.

\subsection{Aggregate Recovery and Downstream Disparity}
\label{sec:aggregate}

STRATA is intended for aggregate population-level auditing, not individual classification, so we validate that use directly. Two questions matter: how well the model recovers group \emph{totals}, and how well it recovers a \emph{disparity} in a downstream outcome. For both, the relevant quantity is the summed predicted probability $\widehat{N}_c = \sum_i p_{ic}$ (and probability-weighted outcome averages), not the hard argmax label---hard labels discard probability information that can materially improve aggregate recovery.

\textbf{Group-total recovery (voter holdout, $N=981{,}288$).} Table~\ref{tab:aggregate} compares true class counts with summed-probability and hard-label estimates. Summed probabilities recover every group share to within 0.5 percentage points, whereas hard labels overstate White by 2.99 points and undercount the ``Other'' residual by roughly 90\% (predicting 2{,}509 against 26{,}130 true). The advantage persists within subgroups: the mean per-class share RMSE across the three states is $0.95$ percentage points for summed probabilities versus $2.77$ for hard labels, and across tract-income deciles $0.25$ versus $1.81$. The signed errors (Table~\ref{tab:aggregate}) show the direction that matters for disparity analysis: hard labels systematically over-state the White share ($+2.99$ points) and erase most of the ``Other'' residual ($-2.41$), whereas summed probabilities stay within half a point in every class. Aggregation from probabilities thus produced substantially smaller aggregate errors than hard labels on this holdout---a property directly relevant to the intended aggregate use.

\begin{table}[htbp]
\centering
\caption{Aggregate group-count recovery on the matched voter holdout ($N=981{,}288$). Signed share error ($\widehat{s}_c-s_c$) in percentage points.}
\label{tab:aggregate}
\begin{tabular}{lccccc}
\toprule
Class & True count & Summed-prob & Hard-label & Prob share err (pp) & Hard share err (pp) \\
\midrule
White & 642{,}349 & 639{,}926 & 671{,}682 & $-0.25$ & $+2.99$ \\
Black & 169{,}655 & 169{,}458 & 161{,}145 & $-0.02$ & $-0.87$ \\
Hispanic & 123{,}787 & 122{,}735 & 127{,}430 & $-0.11$ & $+0.37$ \\
Asian & 19{,}367 & 18{,}504 & 18{,}522 & $-0.09$ & $-0.09$ \\
Other & 26{,}130 & 30{,}666 & 2{,}509 & $+0.46$ & $-2.41$ \\
\bottomrule
\end{tabular}
\parbox{\textwidth}{\small \textit{Note:} Summed-prob $=\sum_i p_{ic}$ over the STRATA ensemble's class probabilities; hard-label $=$ count of argmax predictions. Share error is \emph{signed} ($\widehat{s}_c-s_c$) in percentage points: hard labels over-predict White by $2.99$ and under-count ``Other'' by $2.41$, while summed probabilities are within $0.46$ in every class. Mean per-class share RMSE, in percentage points: by state $0.95$ (prob) vs.\ $2.77$ (hard); by tract-income decile $0.25$ (prob) vs.\ $1.81$ (hard).}
\end{table}

\textbf{Illustrative downstream disparity calculation (national PPP person-name cohort).} As an end-to-end check we estimate the White-minus-Black gap in the PPP loan-forgiveness rate ($\text{ForgivenessAmount}/\text{CurrentApprovalAmount}\ge 0.99$) three ways: from self-reported race, from STRATA hard labels, and from STRATA summed probabilities, on the $108{,}560$ person-name records with a forgiveness outcome. For a proxy that emits class probabilities $p_{ic}$, the probability-weighted group mean is
\begin{equation}
\widehat{\mu}_c = \frac{\sum_i p_{ic}\, y_i}{\sum_i p_{ic}}, \qquad \widehat{\Delta} = \widehat{\mu}_{\mathrm{White}} - \widehat{\mu}_{\mathrm{Black}},
\end{equation}
where $y_i$ is the forgiveness indicator; the hard-label estimator replaces $p_{ic}$ with the one-hot argmax. The forgiveness rate is near-ceiling (99.52\%; self-reported White 99.50\%, Black 99.64\%), so the reference gap is small in absolute terms ($-0.143$ percentage points). STRATA probability weighting reproduces this reference gap most closely ($\widehat{\Delta}=-0.149$ pp, 95\% CI $[-0.22, -0.08]$), whereas the STRATA-hard estimator over-states it ($-0.186$, $[-0.27, -0.10]$) and the BISG-hard estimator shrinks it toward zero ($-0.080$, $[-0.16, 0.00]$)---the majority-class-absorption pattern identified by \citet{Xin2026}---against a self-reported reference of $-0.143$ ($[-0.23, -0.06]$). Except for BISG-hard, whose interval reaches zero, the 1{,}000-resample bootstrap intervals exclude zero, so each of those estimates is individually distinguishable from zero. The point estimates display the expected ordering---probability weighting recovers the direction and magnitude of the true gap, while the hard-label proxies distort it---but we do not test the pairwise differences between estimators, so the intervals quantify uncertainty around each estimate rather than establishing that the estimators differ significantly from one another. We nonetheless present this as a single-pipeline demonstration of the mechanism---majority-class absorption biases measured disparities toward the majority---rather than a definitive validation of downstream disparity recovery across settings.

\section{Limitations}

Several limitations bound the interpretation and appropriate use of STRATA.

\textbf{Training population representativeness.} The model is trained on Florida, Georgia, and North Carolina voter-registration records, together with the national PPP loan file. Registered voters are not a representative sample of the general population: they skew older, more settled, and differently composed by race and socioeconomic status than the full adult population, and the three-state voter geography does not capture regional naming and settlement patterns nationally. Performance on populations dissimilar to the training distribution---as partly evidenced by the state-level variation on the PPP data---may degrade.

\textbf{Temporal drift.} Name--race associations shift over time through immigration, intermarriage, and naming-convention change. A model trained on a contemporary voter snapshot may lose calibration when applied to records from substantially different periods, and periodic retraining is advisable.

\textbf{Geocoding error propagation.} STRATA depends on assigning each individual to a census tract. Address standardization and geocoding introduce errors---unmatched addresses, coarse fallback geographies, and boundary misassignment---that propagate into the tract features the model consumes. We do not quantify this upstream error here.

\textbf{Five-class simplification.}\label{sec:taxonomy} Collapsing race and ethnicity into five mutually exclusive analytical race/ethnicity classes (White, Black, Hispanic, Asian/Pacific Islander, Other) discards meaningful heterogeneity. Under our harmonization (Section~3.1), Native Hawaiian and Pacific Islander (NHPI) records are folded into Asian/Pacific Islander and American Indian/Alaska Native (AIAN), multiracial, and unclassified records into ``Other''; neither NHPI nor AIAN is a distinct output class. To quantify how poorly the taxonomy serves these populations, we compute subgroup recall using the fine-grained provenance label (\texttt{race\_detail}, retained per Section~3.1) rather than a separate output class: among records whose provenance is NHPI, only 3.8\% are recovered, and among AIAN records only 3.4\%. These populations are effectively unrecoverable from name and geography with current training sources, and additional, better-labeled data would be required to serve them. This is a substantive fairness limitation: the populations least well served by the taxonomy are among the most marginalized, and STRATA should not be relied upon for analysis centered on them.

\textbf{Geographic performance variation.} National validation shows accuracy ranging from above 0.95 in several states to below 0.75 in Hawaii (0.695), Washington, Nevada, Arizona, and Alaska. The Hawaii result likely reflects, in substantial part, the five-class collapse and the limited representation of Hawaii-specific populations: the state's large Native Hawaiian and Pacific Islander population is folded into Asian/Pacific Islander and its large multiracial population into ``Other,'' both imperfect fits, so low accuracy there is consistent with a measurement-design limitation rather than a modeling failure per se. Users analyzing such populations should treat STRATA outputs with corresponding caution.

\textbf{Aggregate use only.} As emphasized throughout, STRATA produces probabilistic population-level estimates and is not accurate enough for individual-level determinations; the limitations above compound at the individual level.

\section{Conclusion}

This paper introduced STRATA, a deep learning method for race and ethnicity imputation that integrates character-level name features with block-group and census-tract geolocation, developed for fair lending compliance applications where existing proxy methods distort measured disparities. On a held-out voter dataset of 981,288 records, the STRATA ensemble achieved 88.8\% accuracy and the lowest macro OvR FPR (0.0475, macro-averaged) among all methods evaluated, at 100\% conditional coverage among eligible tract-resolved records, reducing the White FPR from 28.0\% (CFPB-used BISG) to 17.8\%. A controlled ablation isolates the geolocation contribution ($+2.8$ points of accuracy; White FPR reduced from 0.235 to 0.184). Some comparators attain lower one-vs-rest FPRs for particular minority predicted classes, but those values must be interpreted alongside class recall, precision, coverage, and input requirements. STRATA attains the lowest White FPR while maintaining complete conditional coverage and comparatively strong recall for the principal reported classes. A geographically national held-out PPP test provides a geographically broad in-source evaluation at scale on a distinct applicant population of self-identified small-business borrowers (88.5\% accuracy, 9.81\% White FPR, STRATA ensemble, person-name cohort).

A companion paper applies STRATA to 2.26 million NYC residential deed transactions (1966--2025) to decompose homeownership transfer patterns by race \citep{ChalavadiPastorLeitch2026}. The framing above also clarifies the role of STRATA within a larger measurement pipeline. \citet{Xin2026} show that proxy misclassification distorts regression-based disparity estimates through a confusion-matrix mixing of group effects, and identify, as an open direction, audit methods that explicitly model or correct the correlated proxy errors arising from geographic and socioeconomic structure. STRATA addresses the upstream half of this problem by reducing the majority-class absorption (White FPR) that drives the shrinkage they characterize. A companion uncertainty quantification tool (CREST) is described in a separate paper \citep{CRESTinprep}.

The authors caution regarding the use of these predictive models. While STRATA achieves higher accuracy than existing methods, these models are not sufficiently accurate to aid in decisions regarding any one individual record. Using these methods in a transactional process to identify race for individual decision-making would be reckless. However, these models can be reasonably applied post-transaction on an aggregated population to assess the fairness and equity of the transactional process against a qualified benchmark---the use case for which STRATA was designed.

\impact{STRATA is designed to improve the measurement of demographic disparities where self-reported race and ethnicity are missing, a setting that arises throughout fair-lending, fair-housing, and civil-rights enforcement. The intended positive impact is a reduction in the socioeconomically correlated misclassification that causes prevailing proxy methods such as BISG to understate true disparities: by learning a discriminative name--geography interaction rather than applying neighborhood composition as a fixed Bayesian prior, STRATA lowers one important form of majority-class absorption (the White false positive rate) that can contribute to downward bias in measured disparities, and---as the aggregate-recovery and disparity analyses of Section~\ref{sec:aggregate} indicate---probability-based aggregation from STRATA recovers group totals more accurately than STRATA hard-label aggregation and reproduces the illustrative PPP outcome gap more closely than the hard-label estimates examined. More accurate aggregate measurement can help regulated institutions and regulators identify fair-lending exposure that current methods miss.

These same capabilities carry risks that warrant explicit limits. Race and ethnicity imputation can be misused to make or justify decisions about individuals, to target or profile groups, or to manufacture the appearance of compliance while masking discriminatory practice. We emphasize that STRATA's outputs are probabilistic population-level estimates and are not sufficiently accurate for individual-level decisions; using them in any transactional process to assign race to a specific person would be both technically unsound and ethically impermissible. Imputation models can also encode and propagate biases present in their training data (here, voter-registration and Census sources), and accuracy varies across groups and geographies. Appropriate use is confined to post-hoc, aggregate fairness assessment against a qualified benchmark, with transparency about uncertainty and group-specific error rates. We discuss these constraints in the body of the paper and recommend that any deployment be accompanied by documented governance over permitted uses.}

\acks{The authors disclose the following. STRATA is developed by aequumAI LLC, a company founded by an author of this paper; the authors therefore have a commercial interest in the methods described. Portions of the underlying research were initiated while the authors were engaged in race and ethnicity prediction work for the U.S. Social Security Administration. No external grant funding supported the preparation of this manuscript. Any competing interests outside the submitted work are unrelated to its findings.

\textbf{Reproducibility and artifact availability.} The public artifact at \url{https://github.com/tleitch/strata-sagemaker} (archived at Zenodo, concept DOI 10.5281/zenodo.21361277, which resolves to the latest version) supports independent verification of all empirical results reported in this paper. It contains versioned, de-identified prediction files for every reported model and ablation; the evaluation and plotting code; a locked software environment; artifact checksums; and scripts that regenerate every table, figure, confidence interval, and statistical test. The repository also includes PPP cohort-construction and evaluation scripts and a synthetic test dataset. No names, addresses, loan identifiers, or voter-file identifiers are distributed.

\emph{Data availability, licensing, and maintenance.} The released artifact is hosted at the GitHub repository above and archived on Zenodo (concept DOI 10.5281/zenodo.21361277); reviewers and readers can view and download it at those URLs. Code (evaluation, plotting, and cohort-construction scripts) is released under Apache-2.0 and the released data artifacts (de-identified prediction files and documentation) under CC-BY-4.0; the third-party ZRP and Argyle--Barber prediction files are redistributed as evaluation inputs subject to their respective source licenses. The authors confirm they have the right to release these de-identified materials and accept responsibility in the event of any rights violation. The repository is the maintained point of record and will receive corrections as versioned releases, each archived under the same Zenodo concept DOI; the raw voter-registration and PPP source records are obtained from their original public providers and are not redistributed here.

This release does not enable independent retraining of STRATA. The proprietary training implementation and trained weights are withheld for commercial reasons, and the merged voter-file training corpus is not redistributed because it contains individually race-labeled and geocoded records. During review, no-cost access is provided to the versioned inference artifact used to generate the paper's predictions, allowing reviewers to score the public PPP benchmark and synthetic records. The same artifact is available for commercial deployment through the AWS Marketplace listing ``aequumAI STRATA'' (\url{https://aws.amazon.com/marketplace/pp/prodview-edxki4swy6hvs}; Amazon SageMaker AI real-time endpoint or Batch Transform) under the aequumAI STRATA End User License Agreement (EULA) v3.0; no-cost review and academic use are governed by the accompanying Academic Use Addendum v1.0. Accordingly, the release provides complete computational reproduction of the reported evaluation from archived predictions and controlled functional reproduction of inference, but not independent reproduction of model training.}

\appendix
\section{Per-state performance on the PPP person-name cohort}\label{app:perstate}
Table~\ref{tab:perstate} reports accuracy, balanced accuracy (mean per-class recall over classes present in the state), and sample size for the STRATA ensemble on the national PPP person-name cohort ($N=124{,}914$; Section~\ref{sec:aggregate} and Section~4.8), for the 37 states with at least 500 cohort records, sorted by accuracy. States with fewer than 500 records are omitted to avoid over-reading small cohorts. The pattern noted in Section~4.8 is visible here: accuracy is highest in lower-diversity states and lowest in more diverse states, while balanced accuracy---which weights the smaller minority classes equally---is more uniform.

\begin{table}[htbp]
\centering
\caption{Per-state STRATA ensemble performance on the PPP person-name cohort ($N=124{,}914$), for states with $N\geq500$, sorted by accuracy. Balanced accuracy is the mean per-class recall over classes present in the state.}
\label{tab:perstate}
\begin{tabular}{lrrr}
\toprule
State & $N$ & Accuracy & Balanced acc. \\
\midrule
SD & 1{,}102 & 0.9737 & 0.5163 \\
ND & 936 & 0.9733 & 0.6860 \\
IA & 3{,}744 & 0.9714 & 0.5638 \\
NE & 1{,}865 & 0.9705 & 0.6306 \\
KS & 2{,}126 & 0.9539 & 0.5950 \\
MO & 2{,}878 & 0.9382 & 0.6179 \\
WI & 1{,}978 & 0.9287 & 0.6245 \\
MN & 2{,}909 & 0.9251 & 0.5535 \\
IL & 12{,}879 & 0.9209 & 0.6438 \\
KY & 1{,}461 & 0.9117 & 0.5978 \\
OH & 4{,}503 & 0.9094 & 0.5739 \\
PA & 2{,}449 & 0.9065 & 0.6633 \\
MI & 3{,}060 & 0.8964 & 0.5275 \\
GA & 9{,}977 & 0.8900 & 0.5629 \\
IN & 2{,}467 & 0.8877 & 0.6036 \\
NY & 4{,}974 & 0.8852 & 0.6932 \\
VA & 1{,}317 & 0.8800 & 0.6588 \\
MD & 1{,}614 & 0.8792 & 0.6509 \\
TX & 10{,}770 & 0.8781 & 0.6920 \\
LA & 4{,}288 & 0.8724 & 0.5753 \\
NJ & 1{,}290 & 0.8721 & 0.6871 \\
SC & 1{,}957 & 0.8712 & 0.5668 \\
MA & 1{,}407 & 0.8699 & 0.6613 \\
AR & 1{,}494 & 0.8675 & 0.6475 \\
CT & 694 & 0.8674 & 0.6610 \\
TN & 3{,}667 & 0.8549 & 0.6463 \\
AL & 2{,}833 & 0.8549 & 0.5671 \\
FL & 8{,}137 & 0.8514 & 0.6703 \\
OR & 666 & 0.8438 & 0.5516 \\
NC & 2{,}205 & 0.8422 & 0.6961 \\
CO & 726 & 0.8416 & 0.6148 \\
MS & 2{,}725 & 0.8415 & 0.5635 \\
CA & 10{,}828 & 0.8304 & 0.6554 \\
WA & 1{,}116 & 0.8208 & 0.6765 \\
OK & 1{,}741 & 0.8099 & 0.6549 \\
NV & 1{,}543 & 0.8010 & 0.5662 \\
AZ & 1{,}367 & 0.7952 & 0.6354 \\
\bottomrule
\end{tabular}
\end{table}


\begin{thebibliography}{99}

\bibitem[Argyle and Barber(2024)]{Argyle2023}
Argyle, L. P. and Barber, M. (2024). Misclassification and bias in predictions of individual ethnicity from administrative records. \textit{American Political Science Review}, 118(2):1058--1066.

\bibitem[Census Geocoder(2020)]{CensusGeocoder}
Census Geocoder (2020). censusgeo: Census geocoding api. \url{https://geocoding.geo.census.gov/geocoder/}.

\bibitem[Chalavadi et~al.(2026)]{ChalavadiPastorLeitch2026}
Chalavadi, S., Pastor, A., and Leitch, T. (2026). Who is buying New York City's small homes? Working paper, aequumAI LLC, June 28, 2026. \url{https://ssrn.com/abstract=7022879}.

\bibitem[Chalavadi et~al.(in prep)]{CRESTinprep}
Chalavadi, S., Pastor, A., and Leitch, T. (in preparation). CREST: A companion uncertainty quantification tool for race and ethnicity proxy models. Manuscript in preparation, aequumAI LLC.

\bibitem[Elliott et~al.(2008)]{Elliott2008}
Elliott, M. N., Fremont, A. M., Morrison, P. A., Pantoja, P., and Lurie, N. (2008). A new method for estimating race/ethnicity and associated disparities where administrative records lack self-reported race/ethnicity. \textit{Health Services Research}, 43(5p1):1722--1736.

\bibitem[Elliott et~al.(2009)]{Elliott2009}
Elliott, M. N., Morrison, P. A., Fremont, A., McCaffrey, D. F., Pantoja, P., and Lurie, N. (2009). Using the census bureau's surname list to improve estimates of race/ethnicity and associated disparities. \textit{Health Services and Outcomes Research Methodology}, 9(2):69--83.

\bibitem[Fiscella and Fremont(2006)]{Fiscella2006}
Fiscella, K. and Fremont, A. M. (2006). Use of geocoding and surname analysis to estimate race and ethnicity. \textit{Health Services Research}, 41(4p1):1482--1500.

\bibitem[Greenwald et~al.(2024)]{Greenwald2024}
Greenwald, D. L., Howell, S. T., Li, C., and Yimfor, E. (2024). Regulatory arbitrage or random errors? Implications of race prediction algorithms in fair lending analysis. \textit{Journal of Financial Economics}, 157:103857.

\bibitem[Imai and Khanna(2016)]{ImaiKhanna2016}
Imai, K. and Khanna, K. (2016). Improving ecological inference by predicting individual ethnicity from voter registration records. \textit{Political Analysis}, 24(2):263--279.

\bibitem[Imai et~al.(2022)]{ImaiOlivellaRosenman2022}
Imai, K., Olivella, S., and Rosenman, E. T. (2022). Addressing census data problems in race imputation via fully Bayesian Improved Surname Geocoding and name supplements. \textit{Science Advances}, 8(49):eadc9824.

\bibitem[NC State Board of Elections(2022)]{SBE2022}
North Carolina State Board of Elections (2022). Voter registration data. \url{https://www.ncsbe.gov/results-data/voter-registration-data}.

\bibitem[Parasurama(2021)]{Parasurama2021}
Parasurama, P. (2021). raceBERT -- a transformer-based model for predicting race and ethnicity from names. arXiv preprint arXiv:2112.03807.

\bibitem[Sood(2017)]{sood2017}
Sood, G. (2017). Florida voter registration data.

\bibitem[Sood and Laohaprapanon(2018)]{SoodLaohaprapanon2018}
Sood, G. and Laohaprapanon, S. (2018). Predicting race and ethnicity from the sequence of characters in a name. arXiv preprint arXiv:1805.02105.

\bibitem[SURGEO(2022)]{surgeo}
SURGEO (2022). Surgeo documentation. GitHub repository.

\bibitem[Voicu(2018)]{Voicu2018}
Voicu, I. (2018). Using first name information to improve race and ethnicity classification. \textit{Statistics and Public Policy}, 5(1):1--13.

\bibitem[Xie(2021)]{Xie2021}
Xie, F. (2021). Rethnicity: Predicting ethnicity from names. arXiv preprint arXiv:2109.09228.

\bibitem[Xin et~al.(2026)]{Xin2026}
Xin, X., Hooker, G., and Huang, F. (2026). How proxy race distorts regression-based fairness audits. arXiv preprint arXiv:2603.17106.

\bibitem[Zest AI(2022a)]{ZestAIRelease2022}
Zest AI (2022). Uncovering hidden disparities with the zest race predictor. Technical report, Zest AI.

\bibitem[ZestAI(2022b)]{ZestGitHubZRP}
ZestAI (2022). zrp: Zest race predictor. GitHub repository.

\bibitem[ZestAI(2022c)]{ZestAI}
ZestAI (2022). New race prediction model to reduce bias in lending. Press Release.

\end{thebibliography}
\end{document}